# Concerted Rolling and Membrane Penetration Revealed by Atomistic Simulations of Antimicrobial Peptides

Jacob M. Remington, Jonathon B. Ferrell, Jianing Li*.

Department of Chemistry, the University of Vermont, Burlington, VT 05405

*Corresponding authors:
Jianing Li (jianing.li@uvm.edu)

**Abstract**
Short peptides with antimicrobial activity have therapeutic potential for treating bacterial infections. Mechanisms of actions for antimicrobial peptides require binding the biological membrane of their target, which often represents a key mechanistic step. A multitude of data-driven approaches have been developed to predict potential antimicrobial peptide sequences; however, these methods are usually agnostic to the physical interactions between the peptide and the membrane. Towards developing higher throughput screening methodologies, here we use Markov State Modeling and all-atom molecular dynamics simulations to quantify the membrane binding and insertion kinetics of three prototypical and antimicrobial peptides (alpha-helical magainin 2 and PGLa and beta-hairpin tachyplesin 1). By leveraging a set of collective variables that capture the essential physics of the amphiphilic and cationic peptide-membrane interactions we reveal how the slowest kinetic process of membrane insertion is the dynamic rolling of the peptide from a prebound to fully inserted state. These results add critical details to how antimicrobial peptides insert into bacterial membranes.

**Introduction**

Antimicrobial peptides (AMPs) represent a class of short polypeptides with the ability to disturb the growth of bacteria. The presence of AMPs in the innate immune system of plants and animals further suggests their potential for therapeutic applications[1]. Indeed, the peptide sequence space is vast, yet only a relatively small number (~17,000) AMP sequences have been discovered to date[2]. Significant computational efforts have led to predictive models of AMP sequences[34], and these methodologies often include physics-based models of AMPs interactions with model biological features such as membranes[5]. Accurately modeling such AMP-membrane interactions remains a challenging task as membranolytic activity of AMPs occurs over large time- and length-scales with multiple intermediate steps[6]. Here we report on the use of the peptide roll, tilt, and depth coordinates to resolve the initial membrane insertion of three prototypical AMPs, magainin 2 (MAG), PGLa, and Tachyplesin 1 (TAC) (Figure 1). The simulations, totaling 9.6 ms per peptide, were analyzed with Markov state models (MSMs) and revealed that the roll of the amphiphilic AMPs is dynamically coupled with depth of the AMP in the membrane, which together accurately resolve the slowest kinetic process of membrane insertion. These results will guide efforts to screen AMP sequence candidates by providing CVs appropriate for enhanced sampling techniques of AMP insertion, while also reveal critical physical interactions that underly insertion of amphiphilic peptides into biological membranes.

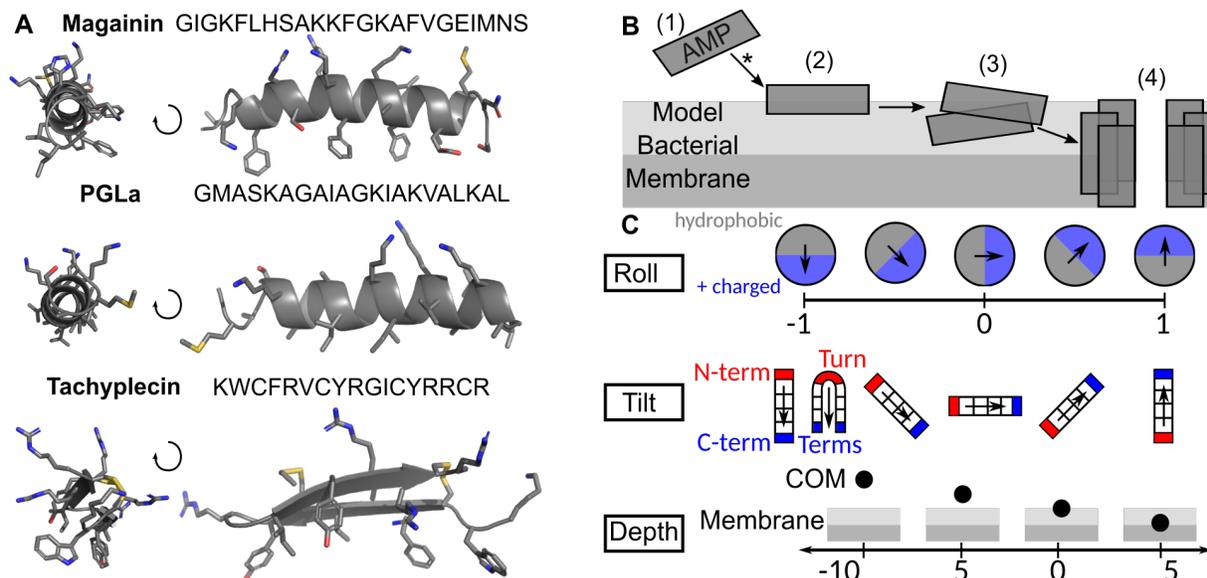

**Figure 1. A** Sequences and structures for the AMPs simulated. **B** Mechanism for amphiphilic AMP activity including the initial AMP insertion (1) to (2) studied here, dimerization (2) to (3), and aggregation to form pores (3) to (4). **C** Proposed collective variables, (roll, tilt, and depth) for AMP insertion. The roll is defined as the z-component of the unit normalized vector pointing from the center of mass of hydrophobic residues to positively charged residues. The tilt is the z-component of the unit normalized vector between the center of masses of the N-term and C-term of helical AMPs and the turn and terminal residues on hairpin AMPs. The depth is the z-component of the center of mass of the CA atoms on the peptide and the upper leaflet P atoms.

The three prototypical AMPs studied in this work have a rich history of experimental data that was considers in the design of our computational approaches. MAG was found using solid state NMR to be highly helical in a membrane inserted state with its helical axis perpendicular to the membrane (tilt = 0) to likely maximize hydrophobic-hydrophobic and hydrophilic-hydrophilic peptide-membrane interactions[7]. As the peptide to lipid molar ratio (P/L) of MAG, increased from 1/30 to 1/10, the surface bound MAG peptides were found to self-assemble, thin the membrane[8], and form toroidal pores using in-plane neutron scattering measurements[9]. These neutron scattering measurements are complemented by the crystallographic observation of toroidal pores with MAG[9]. However, these experimentally resolved mechanistic steps to date do not include the details of the initial insertion event studied herein. NMR measurements have demonstrated PGLa adopts a similar orientation as MAG at low concentration[10] with confirmation from $CF_3$ labels that the roll of the peptide maximizes the hydrophobic-hydrophobic and hydrophilic-hydrophilic interactions provided by the amphiphilic structure of PGLa and the heterogenous structure of lipid bilayers[11]. Increasing the P/L ratio of PGLa changes the orientation of the AMP, with decreases in the tilt angle of the helix with respect to the membrane normal[12,13], finally resulting in fully inserted and vertically oriented PGLa[14]. Interaction of TAC with phospholipid bilayers corresponds with structural changes that increase its amphilicity[15,16] as measured by circular dichroism and fluorescence[17]. Further spectroscopic evidence reveals the Tyr and Trp residues on TAC point towards the interior of the membrane[18]. TAC has also been shown to selectively target POPE/POPG bilayers[19]. While a hairpin, TAC also adopts inserted structures akin to MAG and PGLa with the beta-sheets perpendicular to the membrane normal[20] and hydrophobic residues oriented towards the core of the membrane and charged residues towards the surface[21]. Here we aim to complement what is known about AMP insertion with the dynamic details of the initial membrane bind event that has eluded experiments. More precisely, the experiments suggest a converged orientation for single amphiphilic AMPs, yet they do not provide how this inserted state is formed. Our atomistic molecular dynamics (MD) simulations are designed to explore this key mechanistic step.

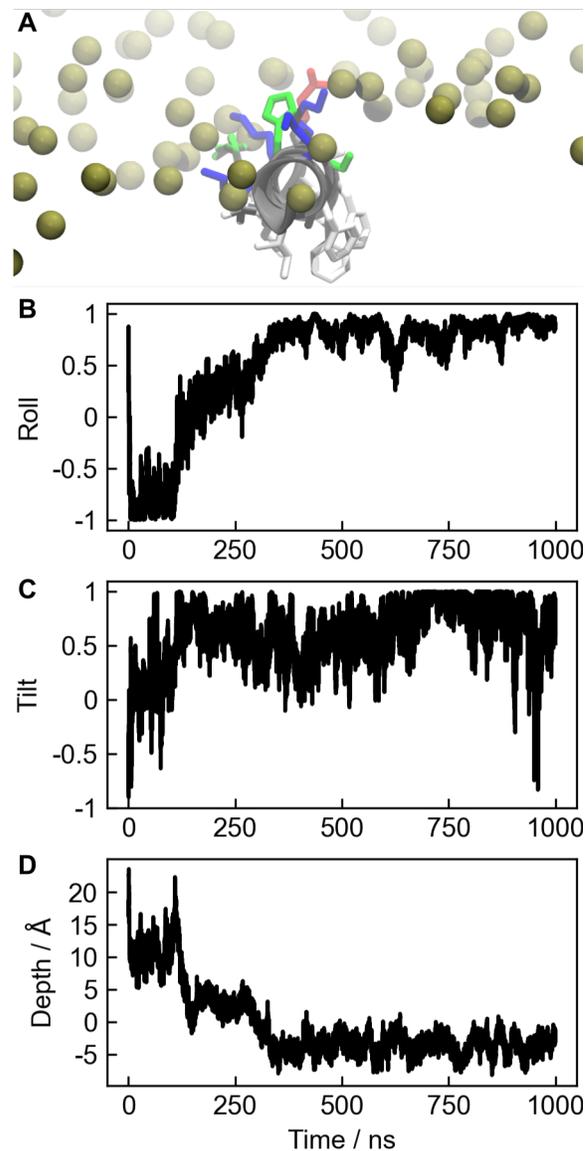

**Figure 2 A** Final snapshot of MAG from a 1 μs simulation initiated above a model bacterial membrane. The peptide sidechains are color-coded by residue type with the non-polar residues white, polar residues green, positively charged residues blue, and negatively charged residues red. The P atoms on the upper leaflet are shown as olive spheres. The positive z-axis is aligned vertically. **B, C,** and **D** Roll, tilt, and depth of MAG demonstrate it initially binds the membrane with a negative roll at times < 100 ns before fulling inserting at 375 ns with a positive roll.

**Results and Discussion**

An all-atom MD trajectory of a single MAG peptide was used as an exploratory step toward revealing the mechanistic details of AMP insertion in a model bacterial inner membrane. During this 1 ms simulation MAG bound and inserted into the bilayer with its helix perpendicular to the bilayer normal (Figure 2A and C). This final structure is reminiscent of the MAG inserted states observed experimentally[7], where the hydrophobic residues have penetrated the hydrophobic core of the bilayer and the positive sidechains form strong interactions with the upper-leaflet phosphate groups. This observation tracks well with the roll of MAG obtaining positive values at the end of the simulation (Figure 2B). However, the MD simulations reveal key mechanistic details about how MAG may form this fully inserted state. First, MAG rapidly bound the membrane in the early 20 ns with the opposite orientation and a roll near -1. This prebound state resulted from the formation of the strong electrostatic interactions between the positive sidechains and the phosphate groups of the bilayer. Notably, the negative roll of this prebound state left the hydrophobic sidechains exposed to water. Over the next 200 ns of this simulation (from times 100 to 375 ns) the peptide reoriented by rolling into the helix to allow the more favorable hydrophobic-hydrophobic interactions and form the fully inserted state. Although just a single trajectory, this simulation suggests rolling plays a critical role in the membrane-insertion mechanism of amphiphilic AMPs and reveals these roll and depth coordinates well distinguish unbound and inserted AMP orientations.

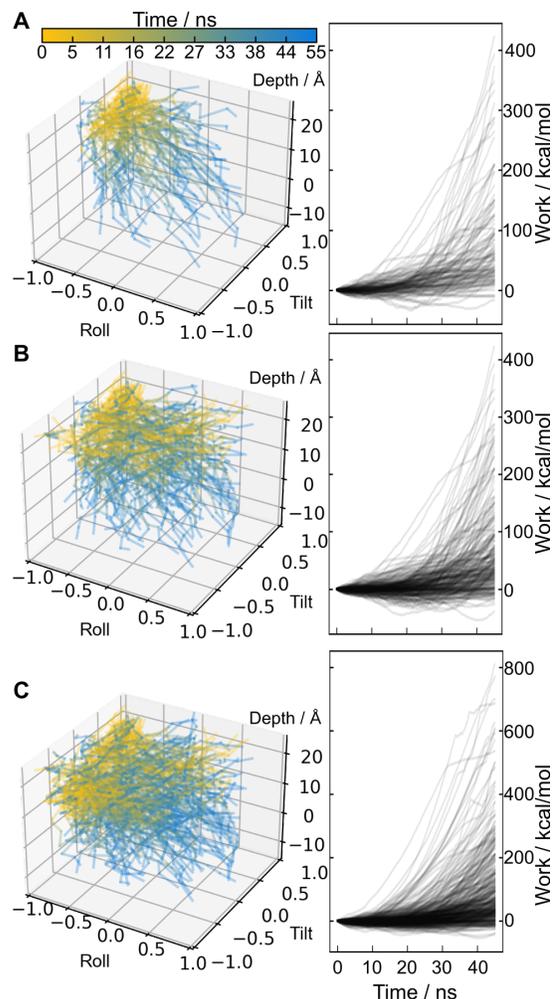

**Figure 3 A** SMD Trajectories of MAG in roll, tilt, and depth space (left) and corresponding work done by the biasing forces during the first 50 ns where moving restraints were applied(right). **B** and **C** The same plots for PGLa and TAC respectively.

To test if the observations from a single AMP binding event hold statistically, large-scale ensembles of AMP simulations were performed and guided by iterative Bayesian estimations of the underlying MSM for membrane insertion of single AMPs. To start, 96 structures spanning the roll, tilt, and depth space were generated using 96 distinct

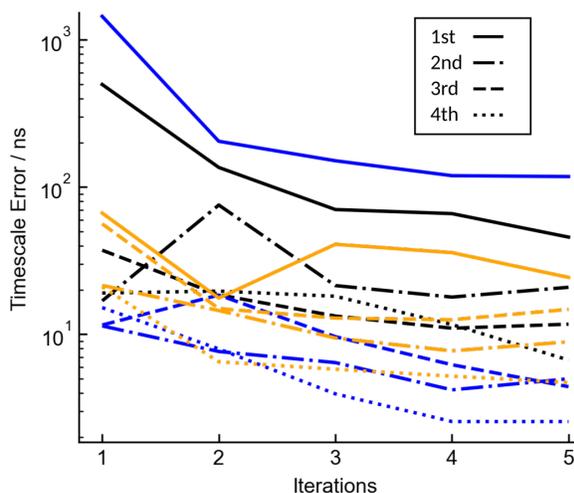

**Figure 4** Decrease in estimated error of the relaxation timescales for Bayesian MSMs of MAG (black), PGLa (blue), and TAC (orange)

50 ns length Steered MD (SMD) simulations (Figure 3). While in principle the work done by the steering force during the SMD trajectories could provide free-energy estimates using the Jarzynski relationship, this approach is notoriously slow to converge and requires multiple simulations of each path. Further, many of the SMD trajectories penetrating deep in the bilayer resulted in large-scale deformations of the bilayer and involved significant work (>100 kcal/mol) to travel through, likely improbable, regions of the coordinate space. Thus, we relaxed each of the resultant structures and proceeded with 20 ns length unbiased simulations. Bayseian MSMs were then estimated from the resultant 1.92 ms of production simulations using a lag time of 15 ns. The relative uncertainties of the eigenvectors corresponding to the four slowest kinetic processes were computed and used as estimates for the regions of the coordinate space where sampling was insufficient. We hypothesized that restarting new simulations from these regions would enhance the overall convergence of the MSM by focusing simulation efforts. Indeed, during four additional iterations of 96 distinct 20 ns simulations we observed a decrease in the error of the four slowest kinetic processes for each peptide (Figure 4). Further, in most cases changes between the 4$^{th}$ and 5$^{th}$ iteration are within the expected error from 4$^{th}$ iteration, signifying convergence of the protocol (Figure 5).

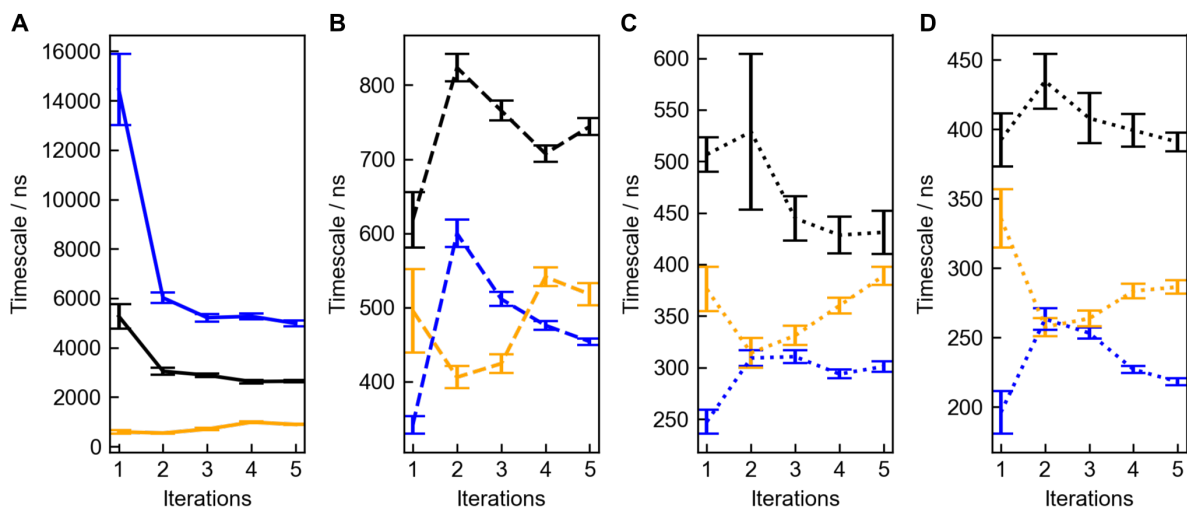

**Figure 5 A** The slowest estimated timescale for Bayesian MSMs of MAG (black), PGLa (blue), and TAC (orange) after successive iterations of the ensemble simulation protocol. **B, C,** and **D** The next three slowest timescales of the MSMs.

The converged MSMs using the five iterations of each simulated AMP enabled the calculation of three-dimensional (3D) free energy surfaces on the roll, tilt, and depth coordinates using the probabilities obtained from the equilibrium eigenvectors. Here, the 3D free energy surfaces are visualized intuitively using depth-wise slices. These slices reveal how the energy landscape along the roll and tilt coordinates change as the peptide inserts into the membrane (Figure 6). Overall, the most favorable roll value of the peptide changes from negative to positive values as the AMP inserts into the membrane. This supports the original observations from the exploratory simulation, in that there are both prebound and inserted states of the AMP in the membrane. In the prebound state, hydrophobic residues are exposed to the solvent while in the inserted state, hydrophobic residues are embedded in the hydrophobic core of the bilayer. The minimum energy insertion pathways computed using structures with roll, tilt, and depth values closest to the depth-wise slice minima are shown in Figure 7 and support this assignment of residue-membrane interactions. With the converged MSM, the transition pathway between these two states is apparent and demonstrates a smooth coupling between the roll and depth coordinates among all three peptides (Figure 6). The other trend observed for all three peptides is that the slowest kinetic process is qualitatively the same among all three peptides and corresponds to transitions between the prebound and inserted states (Figure 6). This later finding follows from the location of the minima and maximum of the first non-equilibrium eigenvector closely matching the locations of these two states in the roll, tilt, and depth coordinates (Figure 6). Given this kinetic event is shared among these three prototypical AMPs it can be hypothesized that the phenomena may be more general among amphiphilic AMPs.

Despite the overall similarities, significant differences between MSMs of the three peptides are apparent. Focusing on MAG, two energy minima are observed; one above (depth, D = -7.5 Å) and one below (D = 5.0 Å) the membrane surface defined by the upper-leaflet P atoms (Figure 6A). For MAG the difference in free energy between the two states is less than RT (~0.6 kcal/mol) with the inserted state slightly stabilized (~0.1 kcal/mol). However, for PGLa the inserted state is stabilized much more (~1 kcal/mol). This difference between PGLa and MAG could be due to differences in the sequence including the presence of a negatively charged residue on MAG or the lack of aromatic hydrophobic residues in PGLa that may facilitate insertion. Furthermore, the slowest timescales for the two helical AMPs, PGL and MAG, were 5 ms and 3 ms respectively (Figure 5A). The slower timescale for PGLa appears to correlate with the larger height of the barrier near D = 0.0 Å in comparison with MAG (Figure 6A and B). On the other hand, the inserted state for TAC was disfavored by approximately 1 kcal/mol relative to the prebound state, although the barrier between the two states is barely noticeable near D = 0.0 Å (Figure 5C). Thus, the rapid exchange between these two states for TAC is excepted and confirmed by the longest timescale for TAC, nearly 1 ms, 3-5 times faster than the helical peptides. This suggests that while the hairpin AMP can rapidly transition between inserted and prebound states, the prebound state predominates. Overall, the slow timescales for AMP insertion highlights the requirements for significant computational efforts when modeling AMP-membrane interactions.

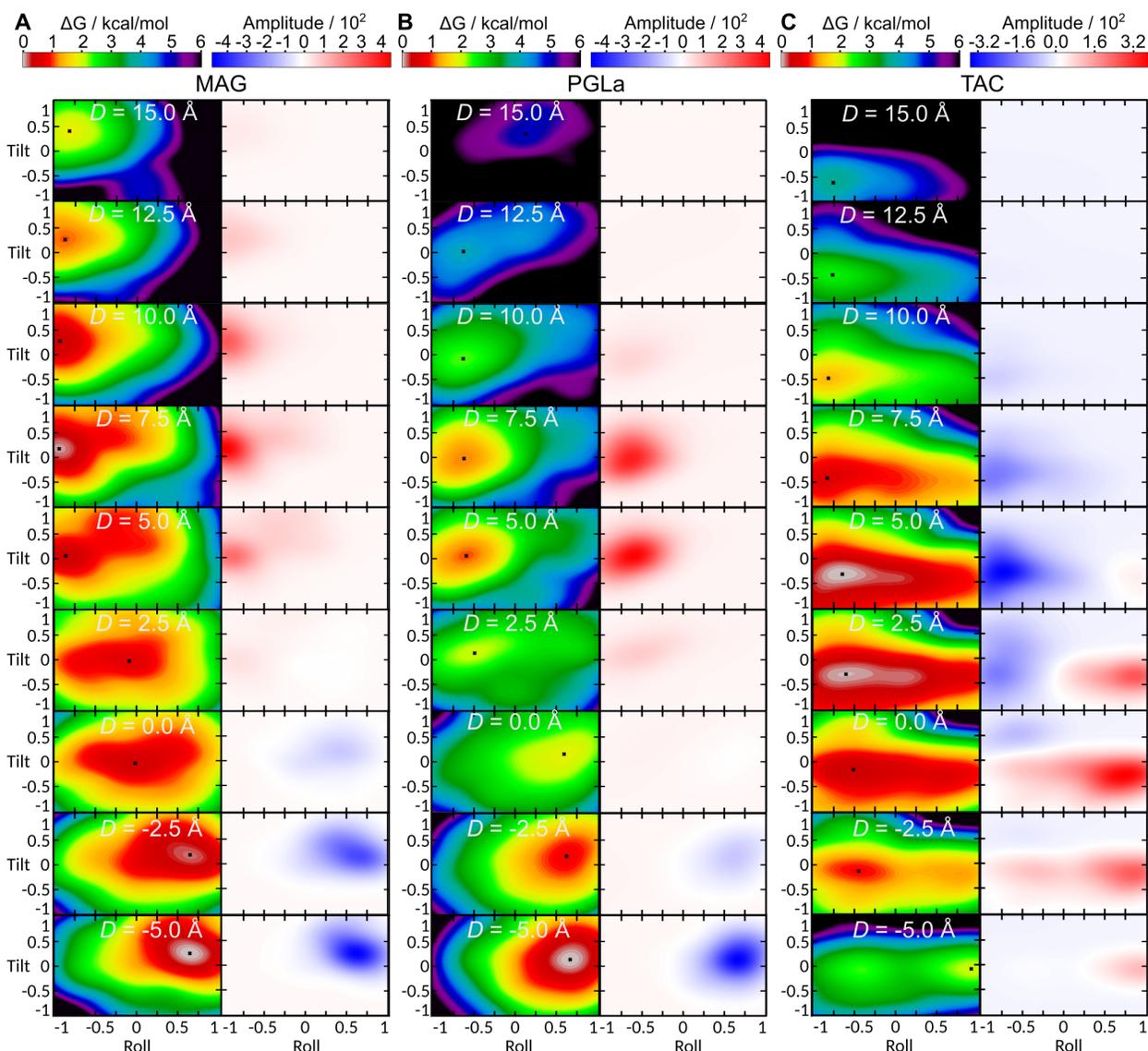

**Figure 6** 3D Free energy surfaces for AMP insertion along the roll, tilt, and depth coordinates are visualized with nine depth-wise slices ranging from 15 to -5 Å (left in each panel). The amplitude of the first non-equilibrium eigenvector of the MSMs in the same coordinate space reveal changes in AMP position and orientation associated with the slowest kinetic event in AMP insertion (right in each panel). **A, B,** and **C** Surfaces for the three peptides MAG, PGL, and TAC are shown respectively. The surfaces are stacked so that the topmost correspond to AMP above the membrane P atoms, and the bottommost have the AMP below the membrane P atoms. The location of the minima and maxima of the non-equilibrium eigenvectors demonstrate that for all three peptides the slowest kinetic event is the rolling of the AMP across the hypothetical membrane surface defined by the P atoms.

**Conclusion**

Ensembles of equilibrium MD simulations guided by iterative MSM estimation revealed the intricate details of AMP insertion into model bacterial membranes. With this approach ~8 ms of aggregate simulation time was required to converge the slowest four timescales in the MSMs of MAG, TAC, and PGL. The converged MSMs demonstrated that the slowest kinetic event in AMP insertion is the rolling of the AMPs amphiphilic moment (defined by the orientation of the hydrophobic and positively charged residues relative to the membrane normal). This rolling is dynamically coupled to the insertion process suggesting that collective variables designed to help sample AMP insertion events should accommodate both coordinates. Our results will likely guide future efforts for more rapidly calculating the free energy changes during AMP insertion into bacterial membrane models. Finally, the finding that all three of the prototypical amphiphilic AMPs had qualitatively similar slowest kinetic processes suggests these results may represent a more general mechanism for insertion of amphiphilic peptides into lipid bilayers.

**Methods**

All-atom models of MAG, TAC, and PGL were prepared using CHARMM-GUI[22]. The exploratory simulation used a 3:1 mixture of POPE and POPG lipids. The MSM simulations used a slightly more accurate to *E. Coli*. Inner membrane model including an additional 5% by mass cardiolipin, POCL[23]. NaCl was used as the salt and counter ions to mimic a salt concentration of 100 mM. A 15 Å buffer was added to the AMP's aligned with a tilt of 0 to define the x and y dimensions of the initial box size. The z-dimension included space to allow AMPs to freely diffuse at least 20 Å above the upper leaflet without interacting (coming within 15 Å) with the image of the lower leaflet. The CHARMM-36m forcefield was used to model all interactions. MD simulations were carried out using the AMBER simulation package[24]. During SMD simulations harmonic restraints were added on root mean squared displacement of peptide CA atoms hold the AMP in their initial structures. Production simulations were conducted at 303 K in the NPgT ensemble using the GPU accelerated version of pmemd. The roll, tilt, and depth coordinates from the aggregate of ensemble simulations were clustered using a uniform space clustering algorithm. Bayesian MSMs using 500 samples of the posterior were estimated using the Pyemma package[25]. At each iteration MSMs were re-estimated, and simulation frames randomly selected from 24 states with the highest relative uncertainty in the first 4 slowest eigenvectors of the MSM. As the amplitudes of eigenvectors are only unique up to a multiplicative scalar, they were first normalized using the 2-norm and for all but the first, $v_0$, of the 500 estimated eigenvectors, $v_i$, the scalar $x$ in {-1,1} that maximized $x*v_i \times v_0$ was used to pick the appropriate $x*v_0$ for computing averages and variances. The 3D free energy surfaces and amplitudes for the 3D non-equilibrium eigenvectors were computed from a kernel density estimation using the probabilities (and amplitudes) as weights. Normalization of the 3D surface were performed before slicing for proper comparison of slices.

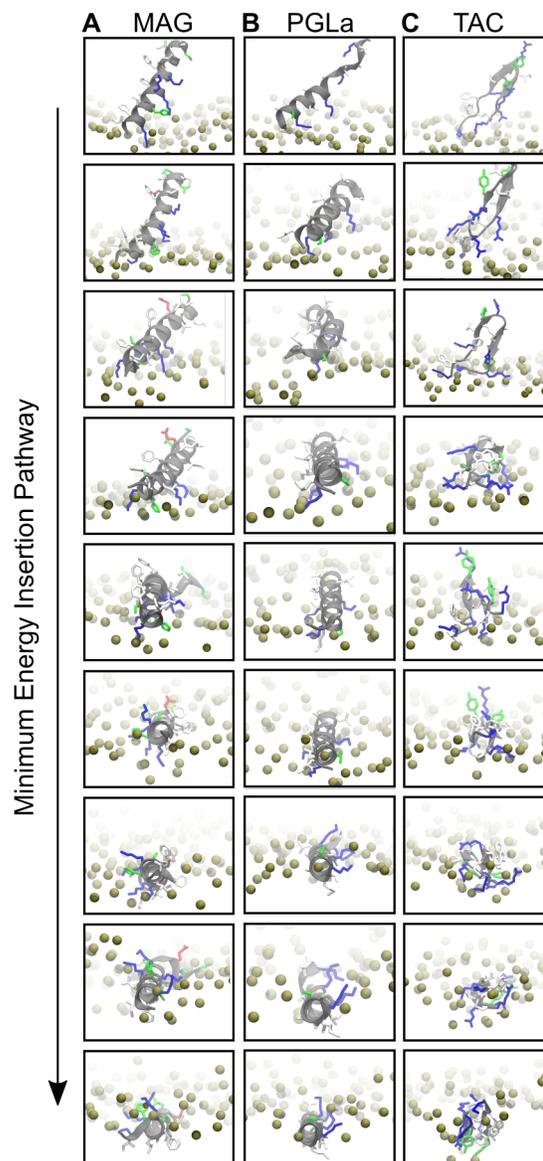

**Figure 7** Minimum energy insertion pathways obtained from the depth-wise slices of the 3D Free energy surfaces in Figure 6 for MAG, PGL, and TAC in **A, B,** and **C** respectively. The peptide sidechains are color-coded by residue type with the non-polar residues white, polar residues green, positively charged residues blue, and negatively charged residues red. The P atoms on the upper leaflet are shown as olive spheres. The positive z-axis is aligned vertically with the page.